\documentclass[12pt]{article}
\usepackage{epsfig}

\newcommand{\NP}{{\em Nucl.\ Phys.\ }}

\newcommand{\PL}{{\em Phys.\ Lett.\ }}
\newcommand{\PR}{{\em Phys.\ Rev.\ }}

\begin{document}
\pagestyle{plain}
\setcounter{page}{1}

\baselineskip16pt

\begin{titlepage}

\begin{flushright}
MIT-CTP-3079\\
NSF-ITP-01-09\\
hep-th/0105156
\end{flushright}
\vspace{13 mm}

\begin{center}

{\Large \bf Gauge Invariance and Tachyon Condensation\\[0.1in]
 in Open String Field Theory\footnote{Based on a lecture presented by
W.\ Taylor at the Strings 2001 conference, Mumbai, India, January 2001.}}
\vspace{6mm}

\end{center}

\vspace{6 mm}

\begin{center}

Ian Ellwood and Washington Taylor\footnote{Current
address:
Institute for Theoretical Physics,
University of California,
Santa Barbara, CA 93106-4030; {\tt  wati@itp.ucsb.edu}}

\vspace{3mm}
{\small \sl Center for Theoretical Physics} \\
{\small \sl MIT, Bldg.  6} \\
{\small \sl Cambridge, MA 02139, U.S.A.} \\
{\small \tt iellwood@mit.edu, wati@mit.edu}\\
\end{center}

\vspace{8 mm}

\begin{abstract}
The gauge invariance of open string field theory is considered from
the point of view of level truncation, and applications to the tachyon
condensation problem are discussed.  We show that the region of
validity of Feynman-Siegel gauge can be accurately determined using
the level truncation method.  We then show that  singularities
previously found in the tachyon effective potential are gauge
artifacts arising from the boundary of the region of validity of
Feynman-Siegel gauge.  The problem of finding the stable vacuum and
tachyon potential without fixing Feynman-Siegel gauge is addressed.
\end{abstract}

\vspace{1cm}
\begin{flushleft}
April 2001
\end{flushleft}
\end{titlepage}
\newpage

\section{Introduction}
The presence of a tachyon in the spectrum of the 26-dimensional open
bosonic string indicates that the perturbative vacuum of this theory
is unstable.  Over a decade ago, Kostelecky and Samuel found evidence
for the existence of a stable vacuum with lower energy, using open
string field theory \cite{ks-open} (see also
\cite{Bardakci-Halpern-tachyon,Bardakci-tachyon}).  At that time,
however, the relevance of D-branes in string theory had not yet been
understood, and the significance of this alternative vacuum was not
appreciated.  In 1999, Sen suggested that the process of tachyon
condensation in the open bosonic string theory should correspond to the
decay of a space-filling  D25-brane, and that open string field theory
should provide a theoretical framework with which it should be
possible to carry out precise calculations describing this decay
process \cite{Sen-universality}.  

In particular, Sen made three conjectures:

\begin{enumerate}
\item[a)] The difference in the action between the unstable vacuum and
the perturbatively stable vacuum should be $\Delta E =VT_{25}$, where
$V$ is the volume of space-time and $T_{25}$ is the
tension of the D25-brane.
\item[b)] Lower-dimensional D$p$-branes should be realized as soliton
configurations of the tachyon and other string fields.
\item[c)] The perturbatively stable vacuum should correspond to the
closed string vacuum.  In particular, there should be no physical open
string excitations around this vacuum.
\end{enumerate}

Sen's conjectures initiated a very active period of research on this
problem.  Conjecture (a) has been verified to a high degree of
precision in level-truncated cubic open string field theory
\cite{Sen-Zwiebach,Moeller-Taylor} and has been shown exactly using
background independent string field theory
\cite{Gerasimov-Shatashvili,kmm,Ghoshal-Sen}.  Conjecture (b) has been
verified for a wide range of single and multiple D$p$-brane
configurations using cubic string field theory
\cite{Harvey:2000tv,deMelloKoch:2000ie,Moeller:2000jy} as well as
background independent SFT (see \cite{Harvey-Komaba} for a review and
further references).  Evidence for conjecture (c) was given in
\cite{Ellwood-Taylor,efhm} using level-truncated open string field
theory.

Until recently most work on this problem using Witten's cubic open
string field theory used the level-truncated approximation to string
field theory in the Feynman-Siegel gauge.  This talk describes some
recent work in which we studied the range of validity of this gauge
choice.  In particular, we have found that singularities in the
tachyon effective potential which appear in Feynman-Siegel gauge are
gauge artifacts associated with the boundary of the region in field
space where Feynman-Siegel gauge is valid.  We also briefly discuss
the problem of solving the tachyon condensation equations in gauges
other than Feynman-Siegel and without any gauge fixing.

Witten's cubic open bosonic string field theory \cite{Witten-SFT} is
defined in terms of a string field
\begin{equation}
\Phi = \phi (p)| \hat{0}; p \rangle+
A_\mu (p) \alpha^\mu_{-1}| \hat{0}; p \rangle + \cdots
\end{equation}
This string field encodes an infinite family of space-time fields,
a space time field
being associated with each ghost number one state in the single string
Fock space.  We write states in the string Fock space in terms of
matter, antighost and ghost creation operators $\alpha^\mu_{-n},
b_{-n}, c_{-m} (n > 0, m \geq 0)$ acting on the vacuum $| \hat{0}
\rangle$.  The string field theory action
\begin{equation}
S =- \frac{1}{2}  \int \Phi \star Q \Phi -\frac{g}{3}  \int \Phi \star
\Phi \star \Phi
\label{eq:SFT-action}
\end{equation}
is defined using the BRST operator $Q$ and a $\star$ product defined
by gluing together the right half of one string with the left half of
another.  This action is
invariant under the infinitesimal gauge transformations
\begin{equation}
\delta \Phi = Q \Lambda + g \left( \Phi\star \Lambda -
\Lambda\star \Phi \right)
\end{equation}
For more background on this cubic open string field
theory, see \cite{Witten-SFT,Gross-Jevicki-12,lpp,Gaberdiel-Zwiebach}.

Previous work on the tachyon condensation problem using cubic open string
field theory has used two simplifications of the full string field
theory defined by (\ref{eq:SFT-action}).   The first simplification is
that
the full string
field theory has been truncated to a finite number of degrees of
freedom by dropping all fields at levels above $L$ (counting the
tachyon as level $L = 0$), and all interactions involving fields whose
levels sum to more than $I$.  This reduction of the degrees of freedom
of the theory is refered to as the  ($L, I$) level truncation of the open
string field theory.

The second simplification of the theory, which is used in most of the
previous work on this subject, is the imposition of the Feynman-Siegel
gauge condition  $b_0 \Phi = 0$.

Using these two simplifications, it has been shown that in the
zero-momentum scalar sector of the string field theory there is a
nontrivial vacuum which appears to be stable under the
level-truncation approximation in the sense that the fields in the
nontrivial vacuum approach fixed values as the level of truncation is
increased.  The energy of this nontrivial vacuum rapidly converges to
the value of $-V T_{25}$ predicted by Sen.  For level truncations up to
(10, 20) we have, for example \cite{ks-open,Sen-Zwiebach,Moeller-Taylor}
\begin{center}
\begin{tabular}{|c|c |l |}
\hline
(L, I) & \# of fields & $E/V T_{25} $\\
\hline
(2, 6) & 3& -0.959 \\
(4, 8) & 10 & -0.986\\
(10, 20) & 252 & -0.999\\
\hline
\end{tabular}
\end{center}

There are several ways in which extending cubic string field theory
computations beyond the Feynman-Siegel gauge may help to shed light on
aspects of the tachyon condensation problem.  For one thing, to
understand the gauge invariance of the theory in the nontrivial
vacuum, it is essential to use the full non-gauge-fixed action.  The
gauge invariant form of the theory is also needed to compute the
cohomology of the shifted BRST operator in the nontrivial vacuum, as
described in \cite{Ellwood-Taylor}.

As we will see later in this talk, another interesting question can be
resolved by considering the range of validity of Feynman-Siegel gauge.
In earlier work using level truncation in Feynman-Siegel gauge, it was
found that the effective potential for the tachyon has branch points
near $\phi \approx -0.25/g$ and $\phi \approx 1.5/g$, in units where the
nontrivial vacuum appears at $\phi \approx 1.09/g$
\cite{ks-open,Moeller-Taylor}.  The locations of these branch points
appear to converge under level truncation to fixed values.  While the
perturbative and nonperturbative vacua both lie on the effective
potential curve between these branch points, the existence of these
branch points makes it unclear what cubic string field theory might
have to say regarding the tachyon potential beyond these branch
points.  In particular, a question of some interest is how the
tachyon potential behaves for large negative values of $\phi$.  In
background independent string field theory, the tachyon potential has
been shown to take the form
$V = (1 + T) e^{-T}$ \cite{Gerasimov-Shatashvili,kmm,Ghoshal-Sen}.
While the field $T$ is related to $\phi$ through a nontrivial field
redefinition, it is clear that this potential is unbounded below as $T
\rightarrow -\infty$, and contains no branch points to the left of the
stable vacuum.  This suggests that the branch point near $\phi \approx
-0.25/g$ found in the cubic theory is not physical.  As we will show,
indeed both of the branch points found in the level-truncated cubic
theory are gauge artifacts.

\section{Gauge Symmetry of Cubic SFT}

We now describe the gauge symmetry of the cubic string
field theory.  We will
restrict attention to scalar fields at zero momentum, which are all
that is relevant for calculations of a Lorentz-invariant vacuum.
The zero-momentum scalar string field can be expanded as
\begin{equation}
\Phi = \sum_i \phi^i| s^{(1)}_i \rangle
\end{equation}
in terms of  ghost number one scalar states in the Fock space.
The action for the component fields $\phi^i$ is a simple cubic
polynomial
\begin{equation}
S = \sum_{i, j} d_{ij} \phi^i \phi^j
+ g\kappa\sum_{i, j, k} t_{ijk}\phi^i \phi^j\phi^k\,
\label{eq:s-action}
\end{equation}
with constant coefficients $d_{ij}, t_{ijk}$, where the constant
$\kappa$ has been chosen so that $t_{111} = 1$.
The scalar
gauge parameters $\mu^a$ are the components of a ghost number zero
string field
\begin{equation}
\Lambda = \sum_{a} \mu^a| s^{(0)}_a \rangle\,.
\end{equation}
The variation of the fields $\phi^i$ under gauge transformations
generated by the $\mu^a$ can be
written as
\begin{equation}
\delta \phi^i = D_{ia} \mu^a + g \kappa T_{ija} \phi^j \mu^a
\label{eq:s-transform}
\end{equation}
with constant coefficients $D_{ia}, T_{ija}$.

We have computed all the coefficients in (\ref{eq:s-action}) up to
level (10, 20), and the coefficients in (\ref{eq:s-transform}) up to
level (8, 16).  We have checked that the action is invariant under the
gauge transformations up to order $g^1$.  At order $g^2$ the gauge
invariance is broken by level truncation, but we have checked that the
gauge invariance is approximately satisfied at this order and improves
as the level of truncation is increased.  The following table lists
the numbers of fields and gauge parameters at each level.
\begin{center}
\begin{tabular}{|c |c |c |}
\hline
Level & Total fields & Gauge DOF\\
\hline
2 & 4 & 1\\
4 & 15 & 5\\
6 & 50 & 19\\
8 & 152 & 61\\
10 & 431 & 179\\
\hline
\end{tabular}
\end{center}

As a simple example of the level-truncated gauge transformations, let
us consider the level (2, 6) truncation.  At this level, the scalar string
field has an expansion
\begin{eqnarray*}
\Phi & = &  \phi| \hat{0} \rangle + B (\alpha_{-1} \cdot \alpha_{-1})|
 \hat{0}
 \rangle 
 + \beta b_{-1} c_{-1}| \hat{0} \rangle + \eta b_{-2} c_{0}|
 \hat{0} \rangle 
\label{eq:s-field}
\end{eqnarray*}
The
gauge parameter at level  two is
\begin{equation}
\Lambda = \mu b_{-2}| \hat{0} \rangle 
\end{equation}
and the gauge transformations are
(in units where $S = -\phi^2/2 + g \kappa \phi^3 + \cdots$)\\
\begin{eqnarray}
\delta \phi & = & g \kappa\mu \left( 
- \frac{16}{9} \phi + \frac{2080}{243} B - \frac{464}{243} \beta +
\frac{128}{81} \eta \right) \nonumber\\
\delta B& = & \frac{\mu}{ 2}  +
g \kappa\mu \left(
 \frac{40}{243} \phi - \frac{9296}{6561} B + \frac{1160}{6561} \beta
-\frac{320}{2187} \eta \right) \label{eq:transform-2}\\
\delta \beta& = &-3 \mu +
g \kappa\mu \left(
-\frac{176}{243} \phi + \frac{22880}{6561} B - \frac{11248}{6561} \beta
-\frac{6016}{6561} \eta \right) \nonumber\\
\delta \eta& = &-\mu +g \kappa\mu \left(
- \frac{224}{81} \phi + \frac{29120}{2187} B + \frac{992}{6561} \beta +
\frac{1792}{729} \eta \right)  \nonumber
\end{eqnarray}
At this level, the action (\ref{eq:s-action}) was computed in
\cite{Rastelli-Zwiebach}.   It can be checked that the action given
there, which agrees with our results, is invariant under this set of
gauge transformations.

\section{Region of validity of Feynman-Siegel gauge}

We can now ask the question of when Feynman-Siegel is a valid gauge
choice.  Near the origin $\phi^i = 0$, the gauge transformations
(\ref{eq:s-transform}) are constant vector fields
\begin{equation}
\delta \phi^i = D_{ia} \mu^a
\end{equation}
Feynman-Siegel gauge sets to zero the fields $\phi^q = 0$ associated
with scalar states in the Fock space annihilated by $c_0$.  It is easy
to see that this gauge choice is good near the origin, in that all
scalar gauge parameters $\mu^b$ associated with scalar states $| s_b
\rangle$ annihilated by $b_0$ give rise to a variation $ \delta
\Phi$ consisting of a part proportional to $c_0 | s_b \rangle$ plus
a part annihilated by $b_0$.  The statement that Feynman-Siegel gauge
is valid locally is equivalent to the condition $\det D_{qb} \neq 0$,
where $q$ ranges over fields associated with ghost number one states
annihilated by $c_0$, and $b$ ranges over fields associated with ghost
number zero states annihilated by $b_0$.  Note that we are using
Feynman-Siegel gauge to fix the set of allowable gauge transformations
as well as to determine which fields are set to vanish.  From this
point of view, a breakdown of Feynman-Siegel gauge can occur either
because a valid infinitesimal gauge transformation produces a vector
tangent to the space of fields in the kernel of $b_0$ or because the
gauge transformation parameters annihilated by $b_0$ no longer span
the space of effective gauge transformations.

At a general point in field space, the gauge transformations are given by
\begin{equation}
\delta \phi^i = M_{ia} \mu^a
\end{equation}
where
\begin{equation}
M_{ia} = D_{ia}+ g \kappa T_{ija} \langle \phi^j \rangle.
\end{equation}
The Feynman-Siegel
gauge choice breaks down when
 \begin{equation}
\det M_{q b} = 0
\label{eq:determinant}
\end{equation}
where $q, b$ are restricted as above to the spaces of states of ghost
number 1/0 annihilated by $c_0/b_0$.  Such a breakdown of gauge fixing
is familiar from nonabelian gauge theories, where Gribov ambiguities
are a well-studied phenomenon.  In string field theory the locus of
points where (\ref{eq:determinant}) is satisfied defines a codimension
one boundary dividing field space into regions where Feynman-Siegel
gauge is locally valid.  The existence of such a boundary means that
not only might several apparently distinct field configurations in
Feynman-Siegel gauge actually be physically equivalent, but there may
also be physical field configurations which have no representative in
Feynman-Siegel gauge.

In the level-truncated theory, $M_{qb}$ is a finite size matrix.
Therefore, we can solve (\ref{eq:determinant}) at a fixed level of
truncation, and we can consider the stability of the locus of points
where this determinant vanishes as we increase the level of
truncation.  We have done this up to level (8, 16) and we find that at
least near the origin the boundary of the region of Feynman-Siegel
gauge validity determined by (\ref{eq:determinant}) seems to be nicely
convergent under level truncation.  

Before discussing the higher level truncations of
(\ref{eq:determinant}), let us consider a simple illustrative example
involving only two fields.  A very simple truncation of the full
string field theory can be defined by dropping all fields except the
level 0 and level 2 fields $\phi$ and $\eta$ defined in
(\ref{eq:s-field}).  Restricted to these two fields, the gauge
transformations are
\begin{eqnarray}
\delta \phi & = & g \kappa \mu
\left(  -\frac{16}{9}  \phi + \frac{128}{81}
 \eta \right)  \label{eq:gauge-restricted}\\
\delta \eta & = &  -\mu+
g \kappa \mu \left(-\frac{224}{81}   \phi + \frac{1792}{729}
\eta \right)\,. \nonumber
\end{eqnarray}
Feynman-Siegel gauge sets $\eta = 0$.
This gauge choice
goes bad when
\begin{equation}
\eta = \delta \eta = 0 \; \;\rightarrow
\; \; \phi = \phi_c =-\frac{1}{g \kappa}  \frac{81}{224} .
\end{equation}
For $\phi <\phi_c$, the 
Feynman-Siegel gauge configurations
with $\eta = 0$ are equivalent under the
restricted gauge transformations (\ref{eq:gauge-restricted}) to
configurations with $\eta = 0, \phi > \phi_c$.  This is analogous to the
appearance of Gribov copies in nonabelian gauge theories.
Furthermore, there is a range of field configurations $(\phi, \eta)$
which have no representatives in Feynman-Siegel gauge. These are the
points with $\phi, \eta< 0$ which lie below the gauge orbit passing
through the point $(\phi, \eta) = (\phi_c, 0)$.  While the full theory
is of course much more complicated after all other fields are
included, this simple two-field example illustrates clearly the
problems which arise when the boundary of Feynman-Siegel gauge
validity is reached.

\begin{figure}
\begin{center}
\epsfig{file=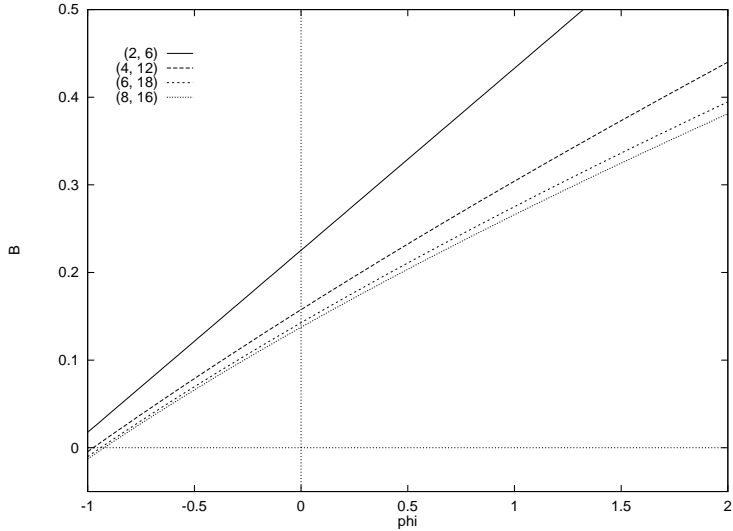,width=10cm}
\end{center}
\caption[x]{\footnotesize Successive approximations to boundary of
region of Feynman-Siegel gauge validity near origin, in units of
$1/3 g\kappa$}
\label{f:pb-region}
\end{figure}

\begin{figure}[htb]
\begin{center}
\epsfig{file=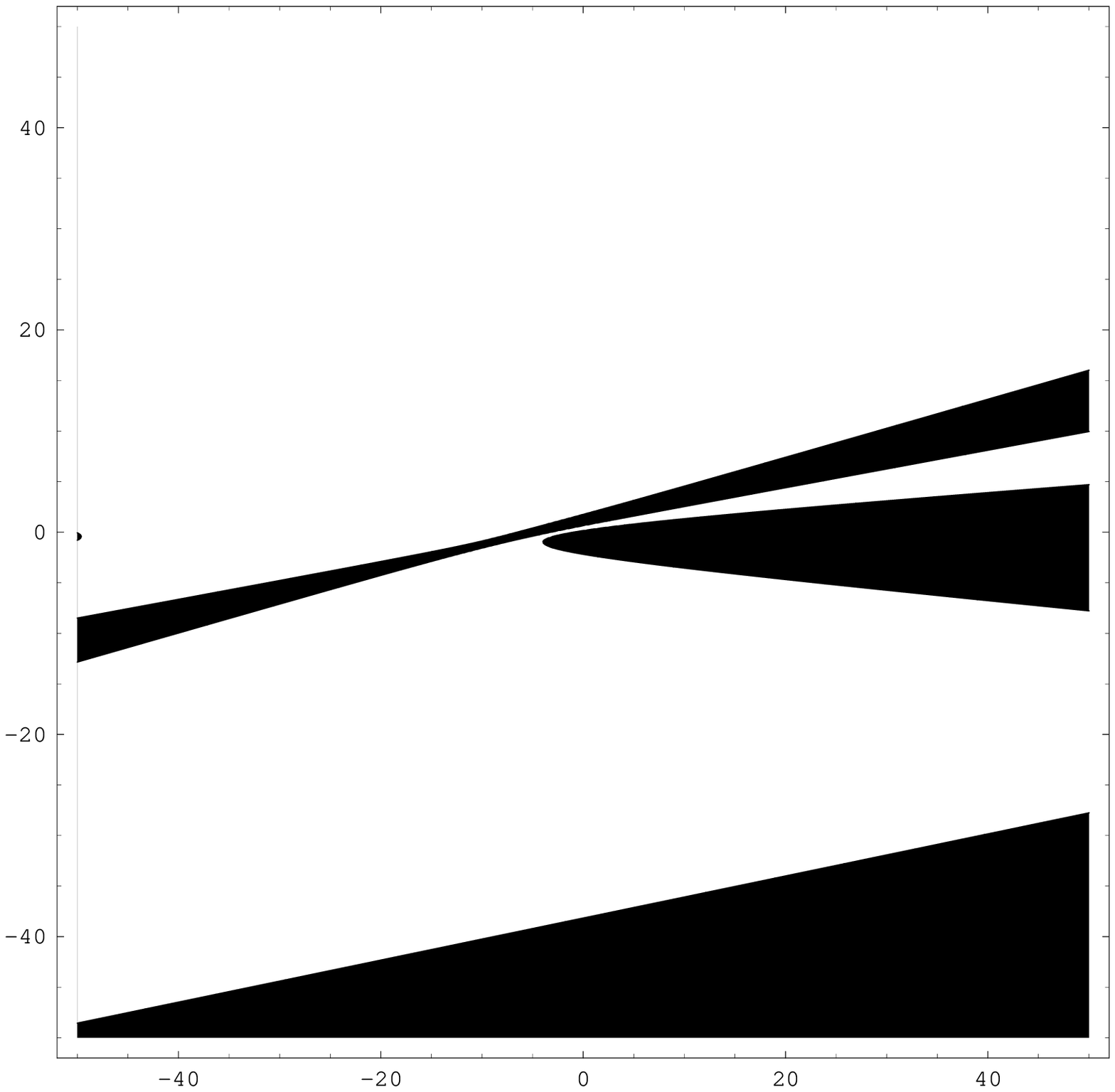,width=6cm}\hspace*{0.3in}
\epsfig{file=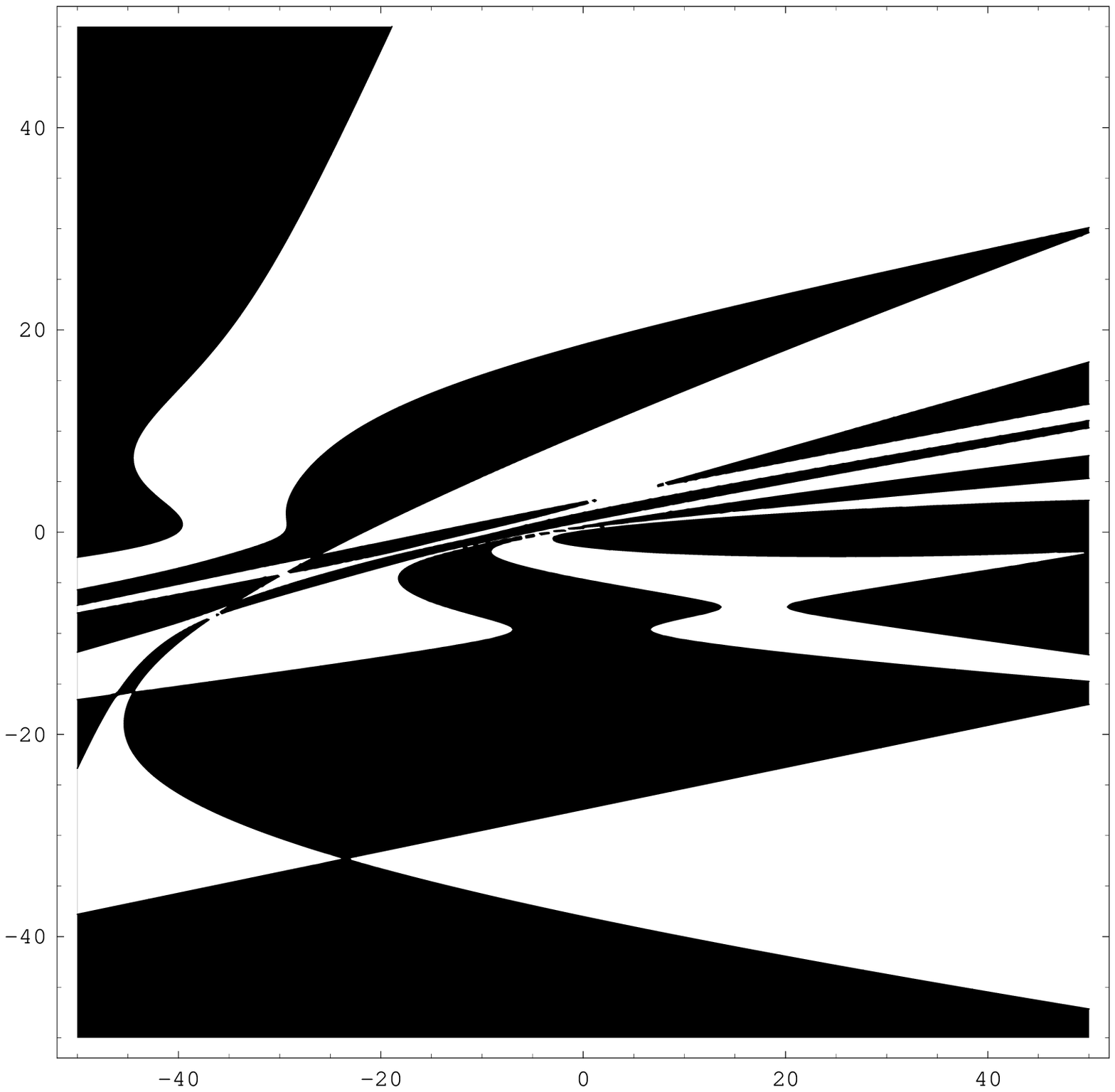,width=6cm}
\end{center}
\begin{center}
\epsfig{file=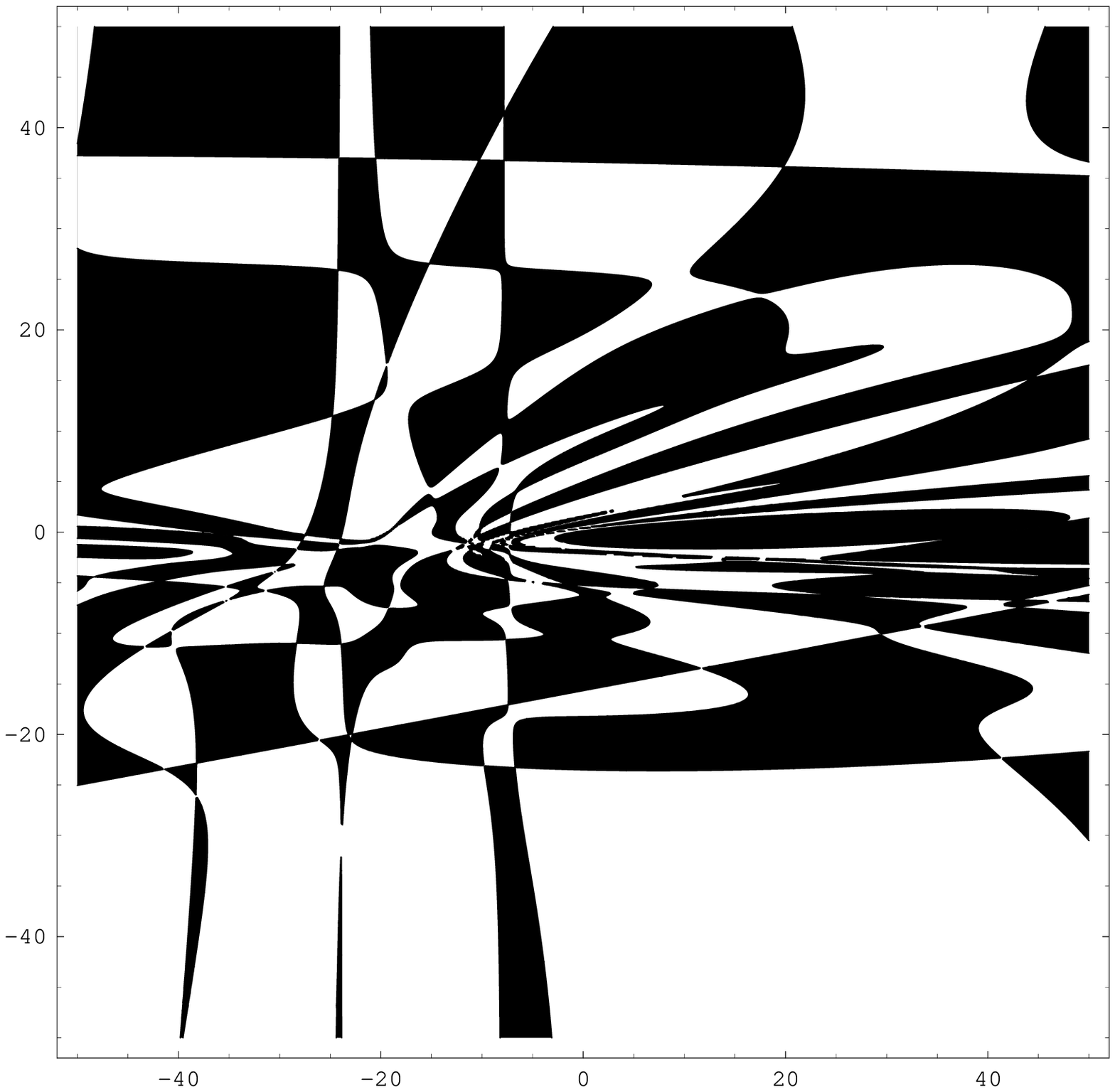,width=6cm}
\end{center}
\caption[x]{\footnotesize Sign[$\det M_{qb}$] at levels (4, 12), (6,
18), and (8, 16)}
\label{f:regions}
\end{figure}

{}From the level two gauge transformations (\ref{eq:transform-2}), we
can see that the level (2, 6) approximation to the determinant condition
(\ref{eq:determinant}) is
\begin{equation}
g \kappa \left(
- \frac{224}{81} \phi + \frac{29120}{2187} B + \frac{992}{6561} \beta 
\right) = 1
\label{eq:}
\end{equation}
This equation defines a plane in the three-dimensional space spanned
by $\phi, B, \beta$.  This plane divides the $\phi-B-\beta$ space into
two regions, in each of which Feynman-Siegel gauge would be locally
valid if the level two gauge transformations were exact.  In
Figure~\ref{f:pb-region} we have graphed the vanishing locus of the
determinant in the $\phi-B$ plane in level 2, 4, 6 and 8 truncations.
It is clear from the figure that the part of this curve near the
origin defines a boundary for the region of Feynman-Siegel gauge
validity which converges fairly well as the level of truncation is
increased.
It seems that the vanishing locus of the determinant becomes less
exactly described by low level truncations as one moves away from the origin.
The sign of the determinant (\ref{eq:determinant}) in the $\phi-B$
plane is shown in Figure~\ref{f:regions} for $-50/3 g \kappa \leq
\phi, B \leq 50/3 g \kappa$ for level 4, 6 and 8 truncations.  The black
elliptical region containing the origin is the region of
Feynman-Siegel gauge validity.  The boundary of this region converges
fairly well near the origin.  Outside this region, the structure of
the sign of the determinant becomes significantly more complicated as
the level of truncation is increased, indicating a very complex
structure for the gauge orbits of the theory.

The results we have summarized here give good evidence that the level
truncation method gives a good systematic approximation scheme for the
location of the boundary of the region of Feynman-Siegel gauge
validity near the origin in field space.  In the remainder of this
talk we discuss briefly a few applications of these results to the
problem of tachyon condensation.

\section{Applications to tachyon condensation}

\subsection{Branch points in tachyon effective potential}

\begin{figure}
\begin{center}
\epsfig{file=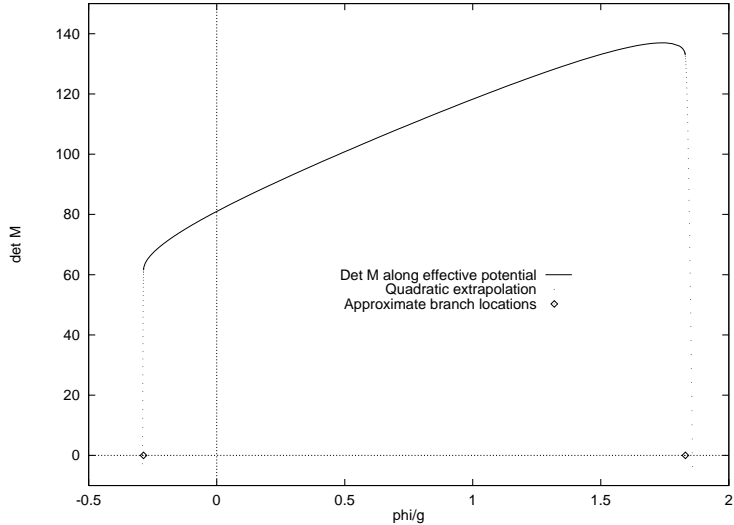,width=10cm}
\end{center}
\caption[x]{\footnotesize $\det M_{qb}$ in level (4, 8) truncation along
tachyon potential.}
\label{f:branch-d}
\end{figure}

As discussed in Section 1, the calculations in \cite{Moeller-Taylor}
of the tachyon effective potential using the level-truncation method
indicated the presence of two branch points in the effective
potential, near $\phi \approx -0.25/g$ and $\phi \approx 1.5/g$.  These
calculations were performed using Feynman-Siegel gauge.  The tachyon
effective potential is computed by choosing a value of $\phi$, solving
the equations of motion for all the other scalar fields, and computing
the energy using these values for the fields.  By following the
trajectory in field space associated with this potential, we have
found that as the tachyon value approaches the points associated with
branch points of the effective potential, the field configuration
given by solving the equations of motion for the other fields
approaches the boundary of the region of Feynman-Siegel gauge
validity.  It is difficult to numerically follow the field trajectory
close to the branch points since numerical methods become unstable in
this region.  By following the trajectory to a point near the branch
point, and extrapolating the trajectory further along a quadratic
curve matching the first and second derivatives of the computed curve,
however, we find that at each level in level truncation, the value of
$\phi$ where the trajectory crosses the Feynman-Siegel validity region
comes within a few percent of the value where the effective potential
encounters a branch point.  An example of such a calculation is shown
in Figure~\ref{f:branch-d}, where the determinant
(\ref{eq:determinant}) is graphed as a function of $\phi$ along the
trajectory of fields giving the effective potential at level $(4,
12)$.  The first branch point at this level is near $\phi \approx
-0.286/g$, and the boundary of Feynman-Siegel gauge validity is
encountered near this point.  Calculating the determinant along the
line of the effective potential to within $0.001/g$ of the branch
point and continuing on a quadratic trajectory, we find that the
boundary of the Feynman-Siegel validity region is encountered at $\phi
\approx -0.290$, within $2\%$ of the branch point location.  Similar
results are found for the second branch point at level (4, 12) and for
both branch points at other levels of truncation.

This analysis gives strong evidence that the branch points in the
tachyon effective potential encountered in \cite{Moeller-Taylor} are
gauge artifacts.  Thus, cubic string field theory seems completely
consistent with background independent string field theory, where the
tachyon potential is unbounded below for sufficiently negative values
of the tachyon field.  We expect that if a method could be found for
using cubic string field theory to compute the tachyon effective
potential below the branch point $\phi \approx -0.25/g$ a similar result
would be found.

\subsection{Extra constraints on Feynman-Siegel gauge solution}

Another interesting result which can be obtained from the level
truncation of the full (non-gauge-fixed) SFT action is an infinite
family of new constraints on the FS gauge nontrivial vacuum state.  It
is now generally believed that there is a well-defined state $\Phi_0$
in the Fock space which is annihilated by $b_0$ and which satisfies
the SFT equation of motion $Q \Phi_0 + g\Phi_0 \star \Phi_0 = 0$.  The
coefficients of the low-level fields in this state were determined to
a high degree of numerical accuracy in
\cite{ks-open,Sen-Zwiebach,Moeller-Taylor} by solving the equations of
motion $\partial S/\partial \phi^b= 0$ for all the fields associated
with states annihilated by $b_0$.  It should also be the case,
however, that the equations of gauge invariance $\partial S/\partial
\phi^q = 0$ associated with states annihilated by $c_0$ should be
satisfied in the nontrivial vacuum.  This gives an infinite family of
additional conditions which should be satisfied by $\Phi_0$.  At any
finite level of truncation, these additional equations of motion will
only be satisfied approximately.  For example, the equation of motion
for the level two field $\eta$ in (\ref{eq:s-field}) vanishes in the
stable vacuum to $92\%$ when only level two fields are considered,
$98\%$ when level four fields are considered, and $99\%$ when level
six fields are included.  Related gauge invariance conditions for the
vacuum were previously considered in
\cite{Hata-Shinohara,Mukhopadhyay-Sen}; other algebraic constraints on
the vacuum were found in
\cite{Zwiebach-constraints,Schnabl-constraints}.  It is interesting to
ask whether all these constraints can be combined to give a more
efficient method for determining the stable vacuum.

\subsection{Finding the vacuum without gauge fixing or in other gauges}

Since we have found that the Feynman-Siegel gauge choice is only valid
locally, it is interesting to ask whether we can find the true vacuum
and/or the tachyon effective potential without gauge fixing.  While in
the complete theory there is an infinite-dimensional gauge orbit of
equivalent locally stable vacua, the breakdown of gauge invariance
caused by level truncation means that even without gauge fixing, the
equations of motion of the level-truncated theory have only a discrete
set of solutions.  In the level (2, 6) truncation, for example, there
are two solutions in the vicinity of the stable vacuum, with energy
densities $E/V T_{25} =$ -0.880 and -1.078 (vs. -0.959 in FS gauge).
(The first of these solutions was also found in
\cite{Rastelli-Zwiebach}).  At level (4, 12), there are at least three
solutions, with $E/V T_{25} = $ -0.927, -0.963, and -1.075
(vs. -0.988 in FS gauge).  As the level of truncation is increased,
the multiplicity of the candidate solutions continues to grow.  While
some solutions approach the correct value, others do not, so without
some further criterion for selecting branches, it does not seem
possible to isolate a good candidate for the vacuum in the
level-truncated, non-gauge-fixed theory.  A unique branch of the
effective potential at each level has the property that it can be
determined by a power-series expansion of the equations of motion
around the unstable vacuum, as was done in \cite{Moeller-Taylor} for
the FS gauge tachyon potential.  Above level 4, however, this branch
encounters a branch point before reaching the stable vacuum.  This
difficulty in solving the theory without gauge fixing clearly arises
from the presence of a continuous family of equivalent vacua in the
full theory.

While solving the theory numerically without gauge fixing does not
seem practical, it is natural to ask whether other gauges may work as
well or better than the Feynman-Siegel gauge for determining the
stable vacuum or the effective tachyon potential.  One simple way of
modifying the Feynman-Siegel gauge choice, for example, is to choose a
different set of fields to vanish at each level from those dictated by
the Feynman-Siegel gauge choice.  As long as these fields are
associated with states in the image of $Q$ in the perturbative vacuum,
this will locally be a valid gauge choice.  As a simple example of
such a different gauge choice, At level two we could choose to fix $B
= 0$ or $\beta = 0$ instead of $\eta = 0$.  More generally, we could
choose to set any linear combination of these fields to vanish which
is not invariant under the linear terms in (\ref{eq:transform-2}).  We
can then take the usual Feynman-Siegel gauge choice for all the
higher-level fields.  Each of these gauge choices defines a new gauge
in which we can perform level truncated calculations to arbitrary
level.  We have tried a variety of gauges of this type.  We find that
in general, these gauges behave rather similarly to Feynman-Siegel
gauge, although the vacuum energy in a generic gauge seems to converge
somewhat more slowly than in Feynman-Siegel gauge.  For example, using
the gauge fixing $B = 0$ for level two fields and FS gauge for all
higher level fields, we find that $E/V T_{25}$ is given by $-0.901,
-0.960, -0.979$ at levels (2, 6), (4, 12) and (6, 18) (compared to
-0.959, -0.988, -0.995 in FS gauge).  By choosing different fields to
vanish at various low levels, we can choose a wide variety of gauges
of this type.  Tuning the coefficients of the linear combinations that
are fixed to zero, we can produce a vacuum approximation at any
particular level of truncation which has an energy density which is
arbitrarily close to the desired value of $-V T_{25}$.  In general,
however, the approach to the vacuum energy is not monotonic, so this
is not a particularly useful way to choose a gauge---even if the
energy is exact at one level, including the next highest level of
fields moves the energy away from the desired value by some small
quantity.

The upshot of this investigation is that while various other gauges
can be chosen which have similar behavior to Feynman-Siegel gauge,
none seem to be particularly better than FS gauge, and generic other
gauge choices seem to lack the property of monotonic convergence of
the vacuum energy which seems empirically to characterize the FS
gauge.  It would be interesting to investigate other more general
gauge choices, such as restricting to fields annihilated by some
particular ghost operator other than $c_0$.  This problem is left to
future work.

\section{Summary}

We have investigated the range of validity of Feynman-Siegel gauge in
Witten's cubic string field theory.  We found that this gauge choice
breaks down outside a fairly small region in field space, and that the
boundary of the region containing the origin in which Feynman-Siegel
gauge is a good gauge choice can be stably computed using the level
truncation method.  We found that branch points appearing in earlier
calculations of the tachyon effective potential are gauge artifacts
arising when the field configuration along the effective potential
leaves the region of validity of FS gauge. We investigated the
possibility of determining the locally stable vacuum and/or the
tachyon effective potential either without gauge fixing or by choosing
a different gauge than Feynman-Siegel gauge, but found no approach
which was substantially better than the Feynman-Siegel gauge-fixed
approach.

\section*{Acknowledgments}

We would like to thank J.\ Harvey, D.\ Kutasov, E.\ Martinec, N.\
Moeller, A.\ Sen and B.\ Zwiebach for helpful discussions in the
course of this work.  WT would like to thank the Institute for
Theoretical Physics in Santa Barbara and the Enrico Fermi Institute
for hospitality during the progress of this work.  The work of IE was
supported in part by a National Science Foundation Graduate Fellowship
and in part by the DOE through contract \#DE-FC02-94ER40818.  The work
of WT was supported in part by the A.\ P.\ Sloan Foundation and in
part by the DOE through contract \#DE-FC02-94ER40818.

\newpage

\bibliographystyle{plain}

\end{document}